\documentclass[12pt,a4paper]{amsart}
\usepackage{fullpage}
\usepackage{graphicx} 
\usepackage{caption}
\usepackage{amsmath,amssymb}
\usepackage{hyperref}
\usepackage{float}
\usepackage{microtype}
\usepackage{color}

\DeclareMathOperator{\area}{area}
\DeclareMathOperator{\gauss}{gauss}
\DeclareMathOperator{\weight}{weight}

\newcommand{\barP}{\overline{P}}

\usepackage{array,multirow,makecell}
\setcellgapes{1pt}
\makegapedcells
\newcolumntype{R}[1]{>{\raggedleft\arraybackslash }b{#1}}
\newcolumntype{L}[1]{>{\raggedright\arraybackslash }b{#1}}
\newcolumntype{C}[1]{>{\centering\arraybackslash }b{#1}}

\date{}
\begin{document}
    \title{Greedy and Local Search Heuristics to Build Area-Optimal Polygons}

    \author{Loïc Crombez, Guilherme D. da Fonseca, and Yan Gerard }

\begingroup
\def\uppercasenonmath#1{} 
\let\MakeUppercase\relax 
\maketitle
\endgroup

	\begin{abstract}
	In this paper, we present our heuristic solutions to the problems of finding the maximum and minimum area polygons with a given set of vertices. Our solutions are based mostly on two simple algorithmic paradigms: greedy method and local search. The greedy heuristic starts with a simple polygon and adds vertices one by one, according to a weight function. A crucial ingredient to obtain good solutions is the choice of an appropriate weight function that avoids long edges. The local search part consists of moving consecutive vertices to another location in the polygonal chain. We also discuss the different implementation techniques that are necessary to reduce the running time.
	\end{abstract}

	\section{Introduction}
\label{sec:introduction}

In this paper, we consider the \emph{optimal area polygonalization} problem, i.e. the problem of finding large and small area simple polygons with a given vertex set. Optimal area polygonalization resembles to the well-known travelling salesman problem, the difference being that the objective function of the former is the area of the computed polygon instead of its perimeter.  This problem has been the subject of the 2019 Geometric Optimization Challenge and is known to be NP-hard for both minimization and maximization~\cite{Fek00}. Exact algorithms are discussed in~\cite{FHKP20} and a recent state of the art is given in~\cite{Dem20}.

In this paper, we describe the algorithm that we developed during the 2019 Geometric Optimization Challenge. Our results gave us the second place, both for the minimization and maximization contests. Throughout, we refer to the \emph{score} of a solution as the area of the polygon divided by the area of the convex hull. The score is a real number between $0$ and $1$, and a lower score is better for the minimization version while a higher score is better for the maximization version.
The scores obtained on the instances of the challenge are in the range $[0.025, 0.352]$ for the minimization problem and in the range $[0.835,0.976]$ for the maximization problem. These two intervals become $[0.110,0.135]$ and $[0.871,0.924]$ if we only consider the uniform instances of at least 100 points where the inputs sets have been randomly generated in a square with a uniform density function. More results are presented in Section \ref{sec:results}.

Our results have been obtained with a relatively simple and fast heuristic coded in Python and executed with pypy3. The heuristic consists of two phases: a greedy heuristic and a subsequent local search optimization. It is surprising that our results are very competitive when compared to the more complex approaches used by the other teams~\cite{Dem20}. The whole source code is available at \href{https://github.com/gfonsecabr/poLYG}{github.com/gfonsecabr/poLYG} and is less than 500 lines long, requiring no external library. While during the challenge we used several different machines, all the running times presented herein have been obtained on a Dell XPS 13-9380 laptop with an Intel i7-10510U CPU and 16GB of RAM running Fedora 32 Linux. The implementation only uses one CPU thread and the other CPU threads were kept mostly idle during the benchmarks.

The paper is organized as follows. Section~\ref{sec:methods} describes the algorithmic approach we used. In Section~\ref{sec:engineering}, we present different techniques implemented to make the code run faster and find better solutions. Section~\ref{sec:results} presents our results. In Section~\ref{sec:outlook}, we discuss some possible improvements. 

	\section{Methods}
\label{sec:methods}

In the next two sections, we describe the two phases of our solution.
For simplicity, we focus only on the \emph{maximum} area polygon. The few changes necessary to solve the \emph{minimization} version are described in Section~\ref{s:min}.

\subsection{Greedy Heuristic} \label{s:greedy}

Let $S$ be the input set with $n$ points. Throughout the execution of the algorithm, we work with a simple polygon $P$ whose set of vertices is a subset of $S$. The set of points in $S$ which are not yet vertices of $P$ is denoted $\bar P$. The current polygon $P$ is initialized with the convex hull of $S$. 

Each greedy step consists of choosing a point $q \in \barP$ and inserting $q$ in the current polygon $P$. We insert $q$ as the intermediary point of an edge $p_1,p_2 \in P$ so that our current polygon $P$ has two new edges $p_1,q$ and $q,p_2$ replacing $p_1,p_2$. 
We preserve at each step the simplicity of $P$ by verifying that the new edges $p_1,q$ and $q,p_2$ do not cross the existing edges of $P$.

\begin{figure}[b]
    \centering
    \includegraphics[width=0.47\textwidth]{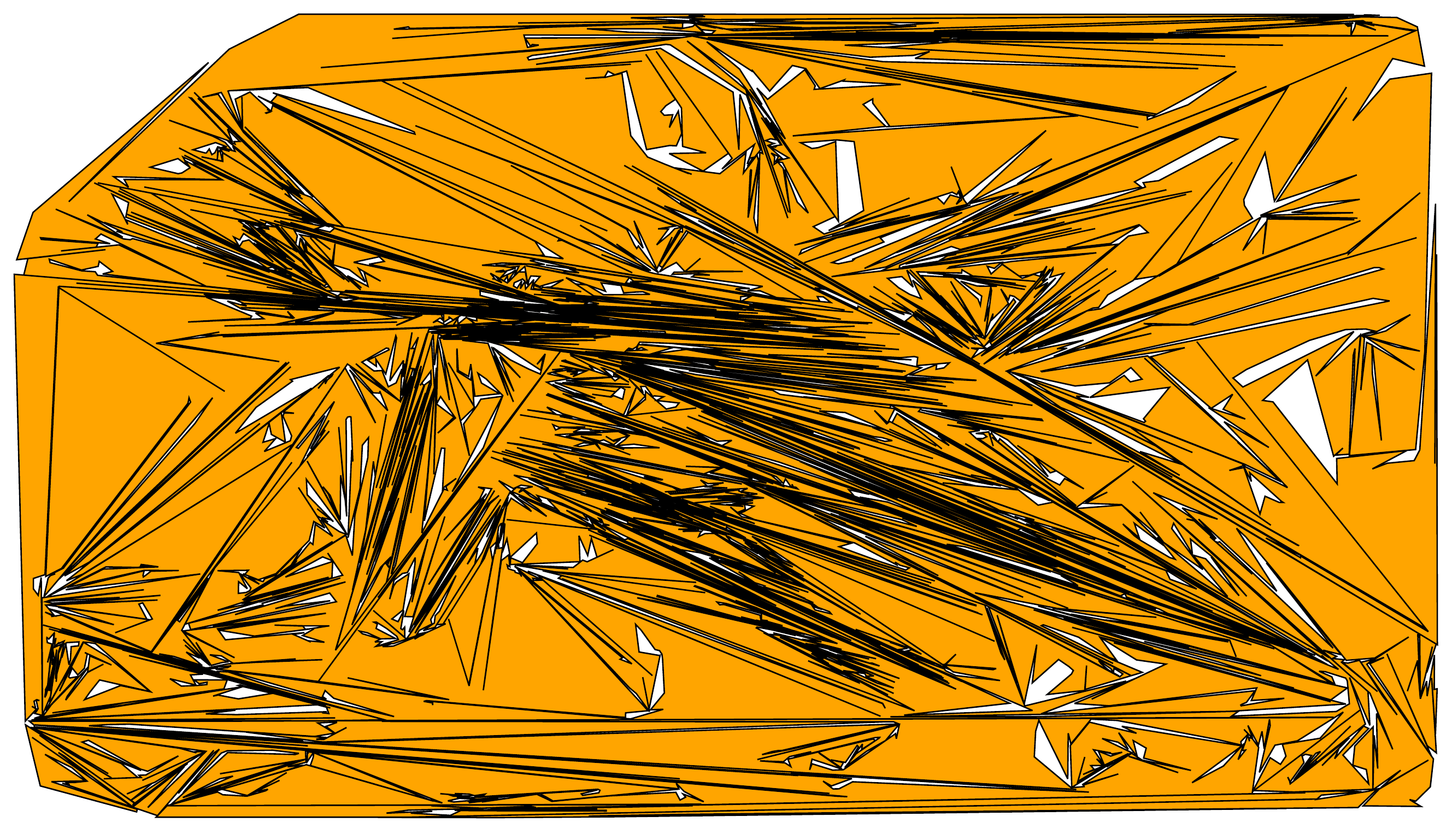} \hspace{.3cm}
    \includegraphics[width=0.47\textwidth]{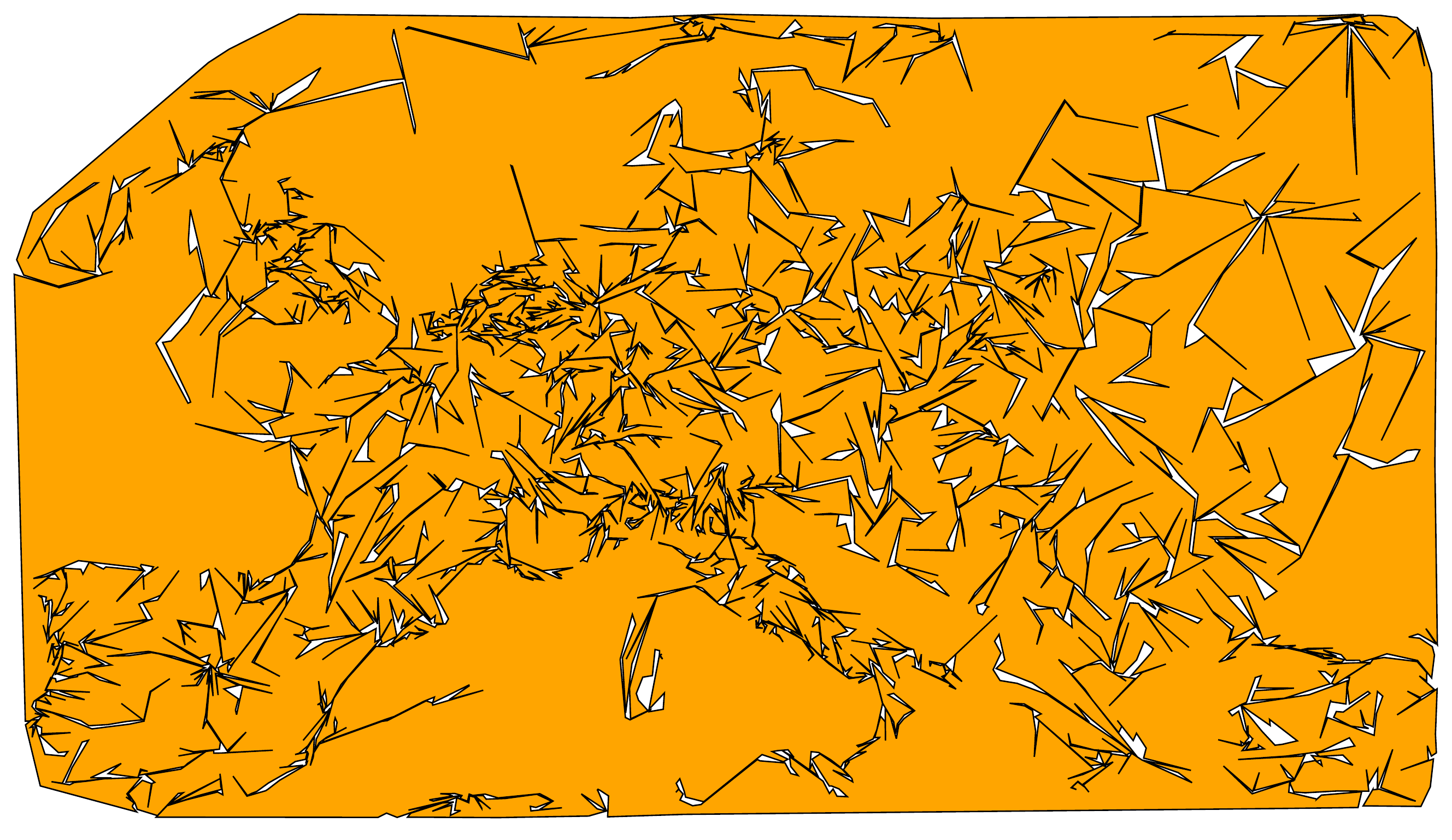}
    \\ \hspace{0cm} (a) \hspace{0.47\textwidth} (b)
    
    \caption{\label{fig:nopen} Two greedy solutions to the maximum area polygon for the \texttt{euro-night-0005000} instance. (a) Each greedy steps chooses the minimum area triangle. (b) Longer edges are penalized with $\alpha = 1/90$. The respective scores are $0.869$ and $0.930$.}
\end{figure}

\begin{figure}
    \centering
    \includegraphics[scale=.7]{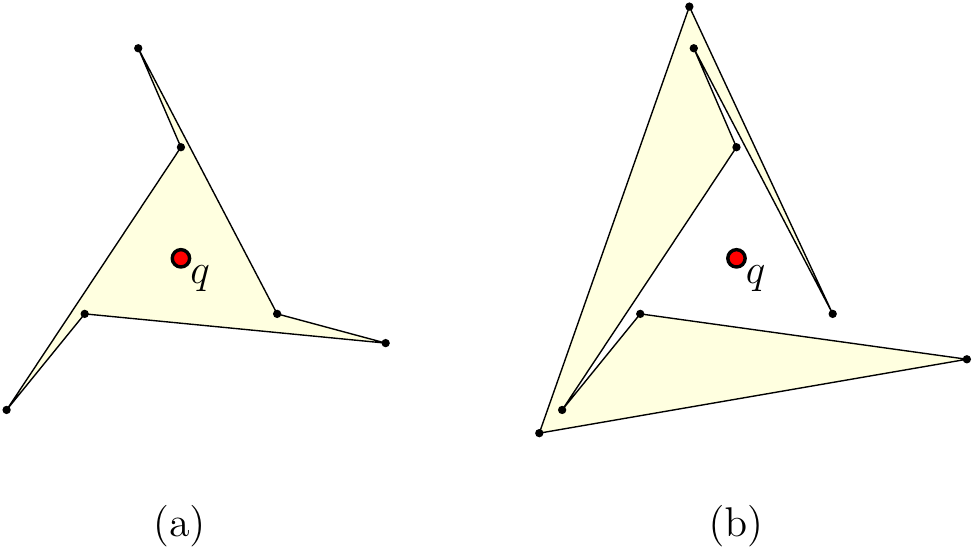}
    \caption{\label{fig:failure} Examples with a red point $q$ (a) inside and (b) outside the polygon such that $q$ cannot be inserted on any of the polygon edges.}
\end{figure}

We repeat our greedy steps until either the set $\barP$ becomes empty or until no point of $\barP$ can possibly be inserted anywhere in $P$ (see Figure~\ref{fig:failure}). In the former case the algorithm successfully finds a solution, and in the latter case it fails. 
Our experiments showed that if the triple $q,p_1,p_2$ is chosen carefully at each step, as we explain in the next paragraphs, then the heuristic hardly ever fails.
Hence, we simply ignored the extremely rare failures and when it happened we used a different value of the parameter $\alpha$ described later on to guide the heuristic.

The choice of the triple $q,p_1,p_2$ at each step is essential to the quality of the solution. A strategy investigated in~\cite{Pee16} is to randomly choose the point $q$ in $\barP$ and then choose the best edge $p_1,p_2$ where to insert $q$. We employ a more exhaustive search.
At each step we choose the triple minimizing $\weight(p_1,p_2,q)$, for a weight function to be described next.
Perhaps, the most natural greedy choice is to minimize the area of the triangle $p_1p_2q$ (denoted $\area(p_1p_2q)$), since the area of $P$ will decrease by exactly $\area(p_1p_2q)$ when $q$ is inserted between $p_1$ and $p_2$. In this case, we say that the weight function is $\weight(p_1,p_2,q)=\area(p_1p_2q)$. This weight function has the property that all points of $\barP$ lie in the interior of $P$. However, this  function leads to very long edges as shown in Figure~\ref{fig:nopen}(a). While long edges may seem like a good choice at short term, they dramatically reduce the search space of potential new triangles, which will hence deteriorate the solution at long term and often make the algorithm fail.

We experimented with several different weight functions in order to obtain better solutions.
The best function that we found by penalizing long edges is
\[\weight(p_1,p_2,q) = \area(p_1p_2q) + \alpha(\|qp_1\| + \|qp_2\| - \|p_1p_2\|),\]
where $\|\cdot\|$ denotes the Euclidean distance and $\alpha$ is a small parameter. The term $\|qp_1\| + \|qp_2\|$ penalizes the creation of long edges, while the term $-\| p_1 p_2\|$ favors breaking existing long edges. Another weight function that gives good results is obtained by replacing the minus sign by a plus sign, and numerous other variations exist. Notice that these functions do not guarantee that all points of $\barP$ lie in the interior of $P$.

The value of $\area(p_1p_2q)$ is a positive number if $q$ is inside $P$ and a negative number otherwise. Using signed areas (negative for clockwise triangles) is important, since some input points in $\barP$ may possibly lie outside $P$. 
We determined that the best values of $\alpha$ are generally in the range $1/150 \leq \alpha \leq 1/50$. Unless otherwise specified, the examples in this paper use $\alpha = 1/90$. Figure~\ref{fig:nopen}(b) shows the improvement obtained by penalizing long edges, while Table~\ref{tab:weight} shows some scores achieved through different weight functions (the local search algorithm is discussed in the next section).

\begin{table}
\begin{tiny}
\begin{tabular}{|L{1.5cm}|C{1.3cm}|C{1.3cm}|C{1.3cm}|C{1.35cm}|C{1.3cm}|C{1.3cm}|C{1.3cm}|C{1.35cm}|}
\hline & \multicolumn{4}{c|}{greedy only} & \multicolumn{4}{c|}{after local search}\\
 & $\alpha=1/10$ & $\alpha=1/30$ & $\alpha=1/90$ & $\alpha=1/270$ & $\alpha=1/10$ & $\alpha=1/30$ & $\alpha=1/90$ & $\alpha=1/270$ \\
\hline
euro-night & 0.848 & 0.885 & 0.893 & 0.886 & 0.896 & 0.911 & 0.912 & 0.908 \\
           & 0.882 & \textbf{0.894} & 0.892 & 0.891 & 0.910 & 0.911 & \textbf{0.920} & 0.916 \\
\hline
paris & 0.833 & 0.871 & \textbf{0.882} & 0.871 & 0.880 & 0.899 & 0.900 & 0.889 \\
      & 0.860 & 0.871 & 0.876 & 0.854 & 0.893 & \textbf{0.902} & 0.898 & 0.877 \\
\hline
stars & 0.833 & 0.858 & \textbf{0.881} & 0.867 & 0.885 & 0.898 & 0.903 & 0.901 \\
      & 0.858 & 0.866 & 0.872 & 0.858 & 0.895 & 0.896 & \textbf{0.903} & 0.889 \\
\hline
us-night & 0.850 & 0.914 & \textbf{0.928} & 0.919 & 0.898 & 0.942 & 0.943 & 0.934 \\
         & 0.907 & 0.920 & 0.924 & 0.920 & 0.933 & 0.939 & 0.943 & \textbf{0.945} \\
\hline
uniform-1 & 0.824 & 0.846 & \textbf{0.859} & 0.841 & 0.865 & 0.872 & \textbf{0.875} & 0.863 \\
          & 0.817 & 0.842 & 0.841 & 0.828 & 0.865 & 0.868 & 0.866 & 0.856 \\
\hline
uniform-2 & 0.795 & 0.838 & \textbf{0.844} & 0.815 & 0.848 & 0.861 & \textbf{0.862} & 0.849 \\
          & 0.816 & 0.830 & 0.837 & 0.813 & 0.860 & 0.858 & 0.857 & 0.846 \\
\hline
\end{tabular}
\end{tiny}
\caption{\label{tab:weight} Scores before and after local search for different values of $\alpha$ and instances with $500$ points. The scores above use the formula $\weight(p_1,p_2,q) = \area(p_1p_2q) + \alpha(\|qp_1\| + \|qp_2\| - \|p_1p_2\|)$ while the ones below use $\weight(p_1,p_2,q) = \area(p_1p_2q) + \alpha(\|qp_1\| + \|qp_2\| + \|p_1p_2\|)$.}

\end{table}

\subsection{Local Search} \label{s:ls}

\begin{figure}[b]
    \centering
    \includegraphics[scale=.7]{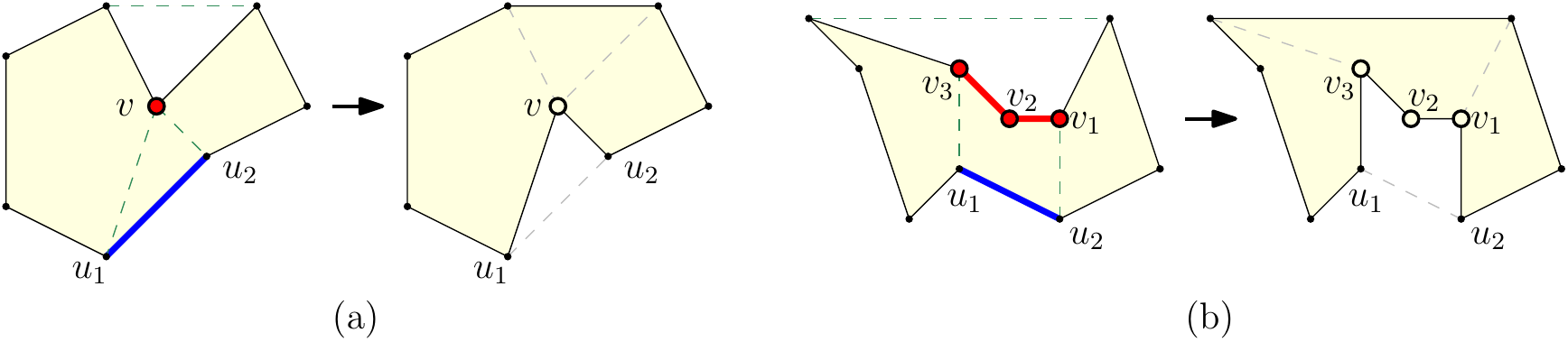}
    \caption{\label{fig:ls} Local search. (a) 
    The blue edge is replaced by a two-edge path that detours to the red point, and the two prior edges incident on the red point are replaced with a single edge. (b) The blue edge is replaced by a detour to utilize the red subpath.
    }
\end{figure}

After obtaining the greedy solution, we perform a second phase to improve the score of the solution, by making local changes to the polygon. The simplest optimization we perform consists of moving one vertex $v$ to another position in the polygonal chain, between the endpoints of an edge $u_1u_2$ (Figure~\ref{fig:ls}(a)), adding the edges $u_1v$ and $vu_2$, while removing the edge $u_1u_2$. A more general version of this procedure consists of moving a path $V=v_1,\ldots,v_k$ (consisting of one or more vertices) together, in reverse order,  between the endpoints of an edge $u_1u_2$ (Figure~\ref{fig:ls}(b)). The order by which we perform the numerous local changes has some impact on the solution. Hence, we chose to generally perform the changes that increase the area the most first. Our algorithm works as follows.

\begin{figure}[t]
    \centering
    \includegraphics[width=0.31\textwidth]{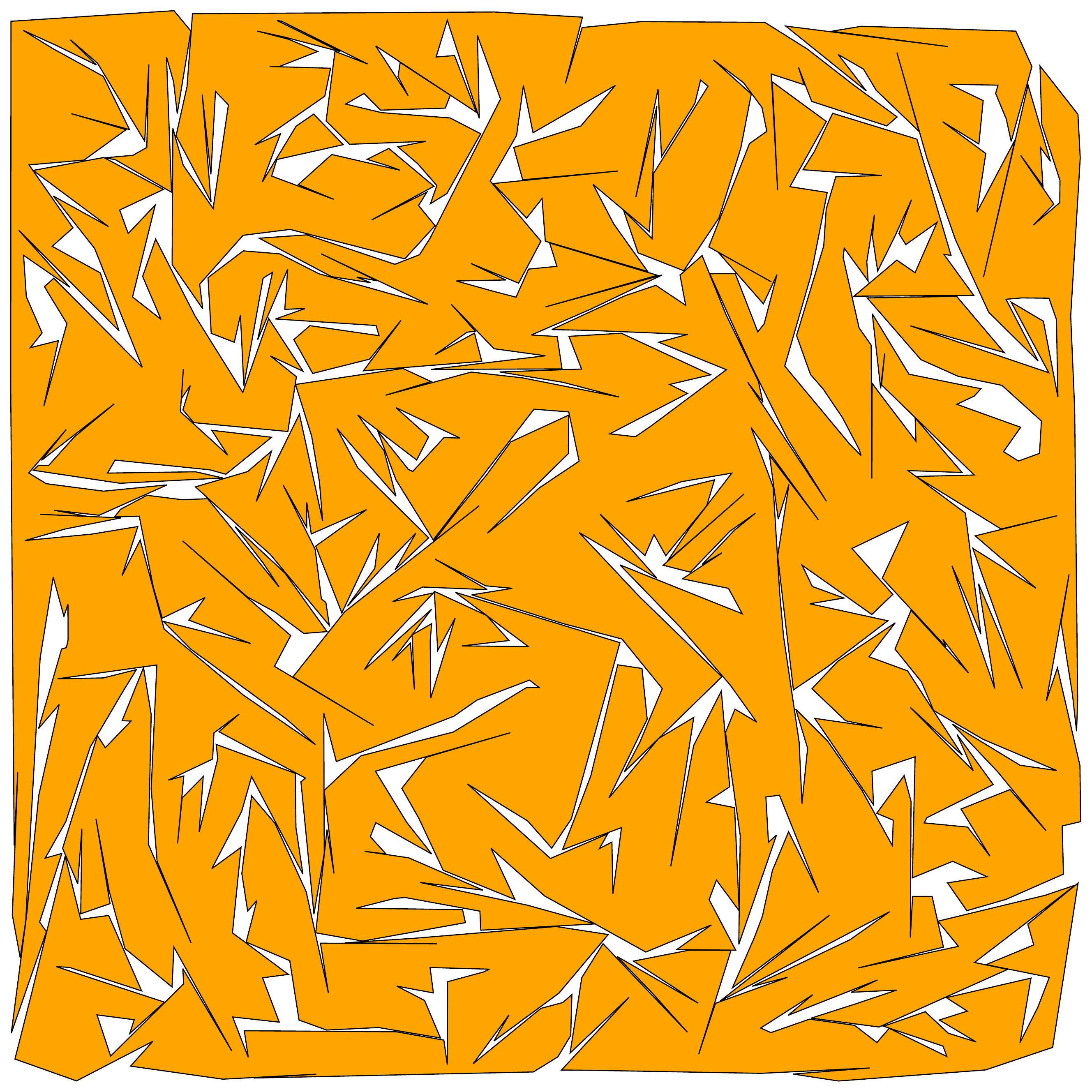} \hspace{.2cm}
    \includegraphics[width=0.31\textwidth]{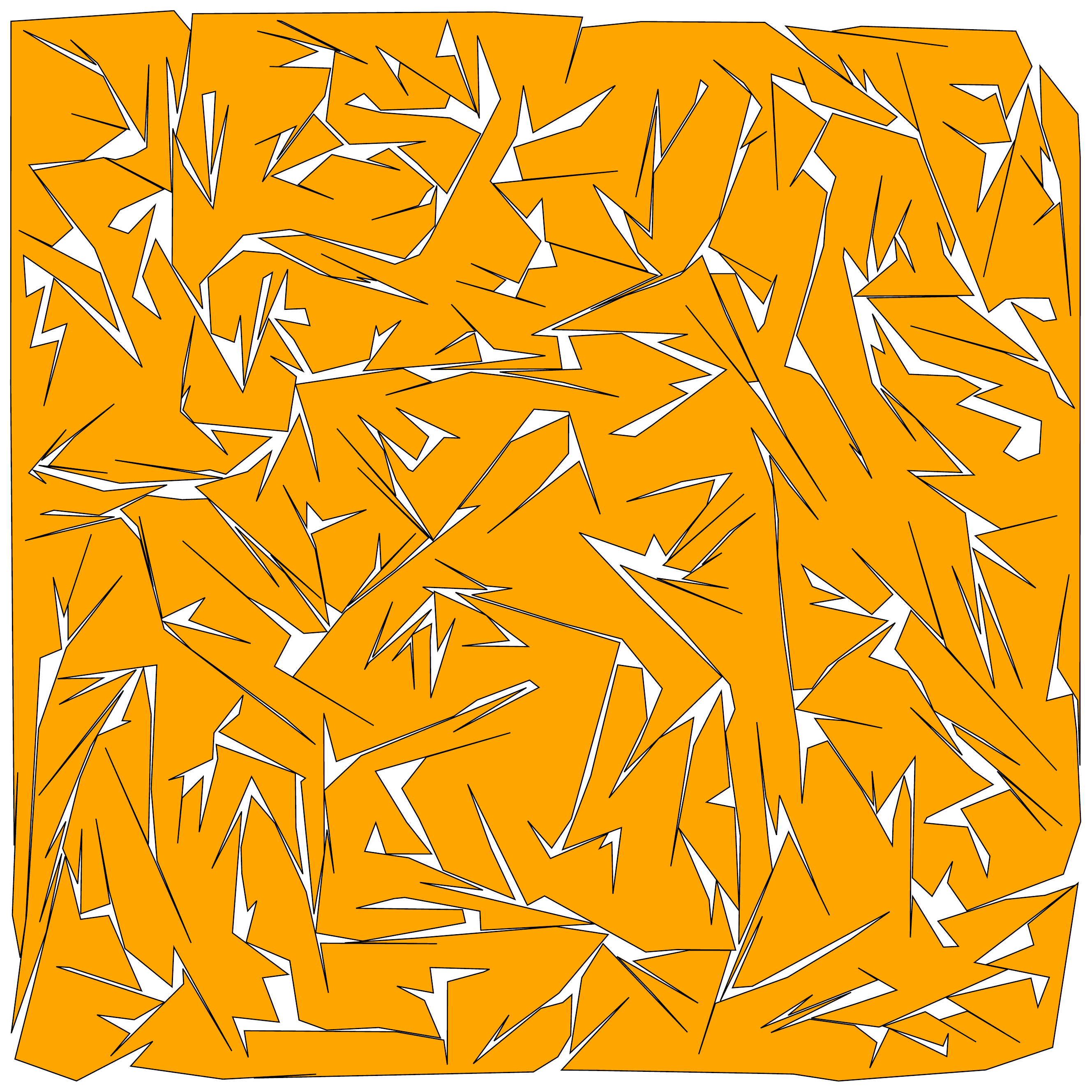} \hspace{.2cm}
    \includegraphics[width=0.31\textwidth]{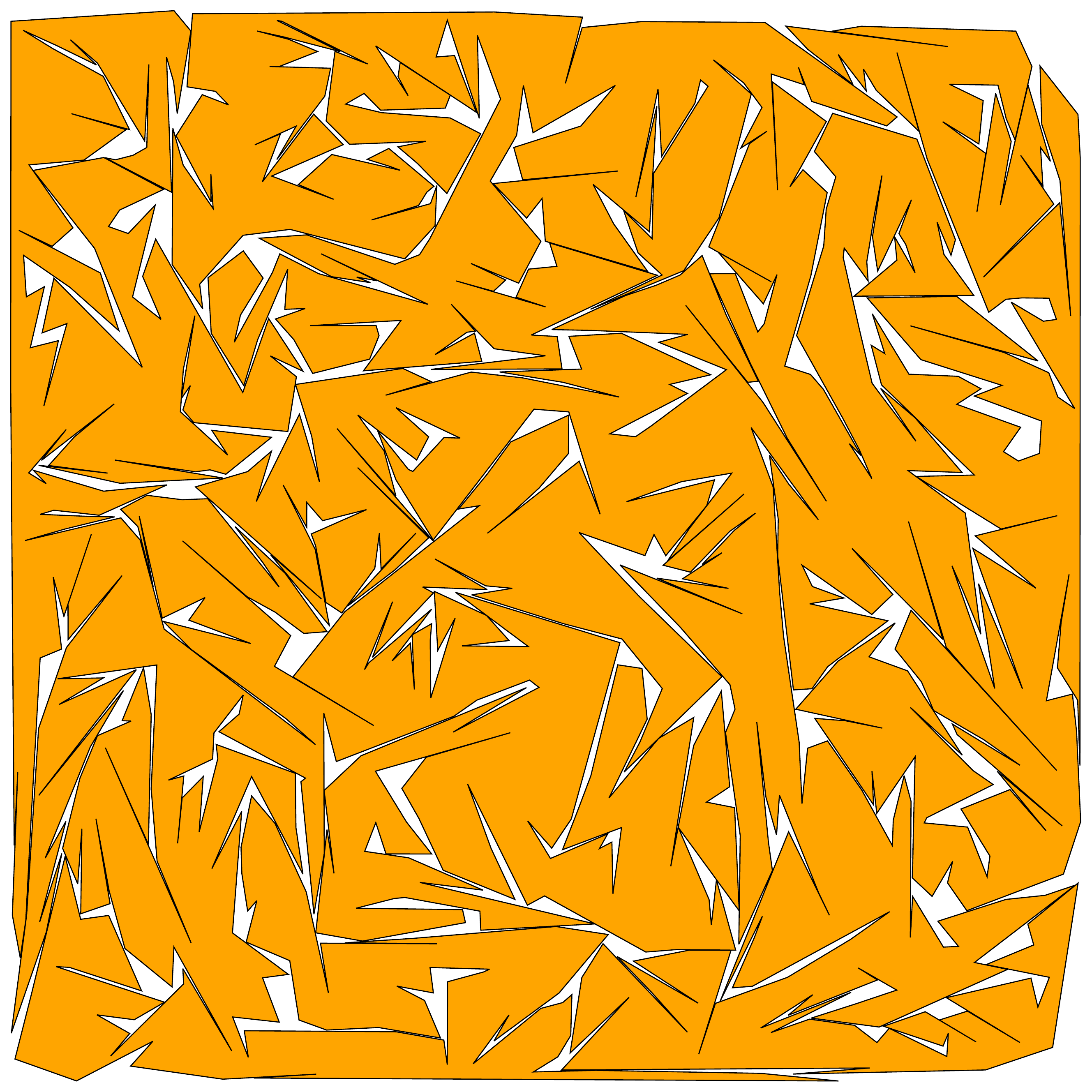}\\
    (a)\hspace{0.31\textwidth}(b)\hspace{0.31\textwidth}(c)
    \caption{\label{fig:grls} (a) Greedy and (b,c) local search solutions to the \texttt{uniform-0001000-1} instance with $\alpha=1/90$. The two local search solutions use $\ell=1$ and $\ell=10$, respectively. The respective scores are $0.842$, $0.863$, and $0.866$.}
\end{figure}

We fix a maximum number of vertices $\ell$ for the path $V$ as a parameter (typically at most $\ell=10$ vertices). Then, we go through every pair of edge $e = u_1u_2$ and path $V$ of length at most $\ell$, testing the following two conditions: moving $V$ to $e$ should (i) increase the area of the polygon and (ii) result in a simple polygon. Every pair $e,V$ that meets these conditions is stored in a list $L$. Then, we sort $L$ from the largest to the smallest area change. Finally, we iterate through $L$ applying the changes in order. However, before applying each change, we need to retest that the modification is still valid (condition (ii)) and beneficial (condition (i)), since previous changes already modified the polygon. We repeat the whole procedure until the improvement becomes negligible (in the test cases presented herein, a score change smaller than $0.001$). An example of the greedy solution as well as two local search optimizations with different values of $\ell$ is presented in Figure~\ref{fig:grls}. To see the differences in this figure, compare the large white areas on the left figure with the corresponding areas on the right figures.

\subsection{Minimization} \label{s:min}

\begin{figure}
    \centering
    \includegraphics[width=0.47\textwidth]{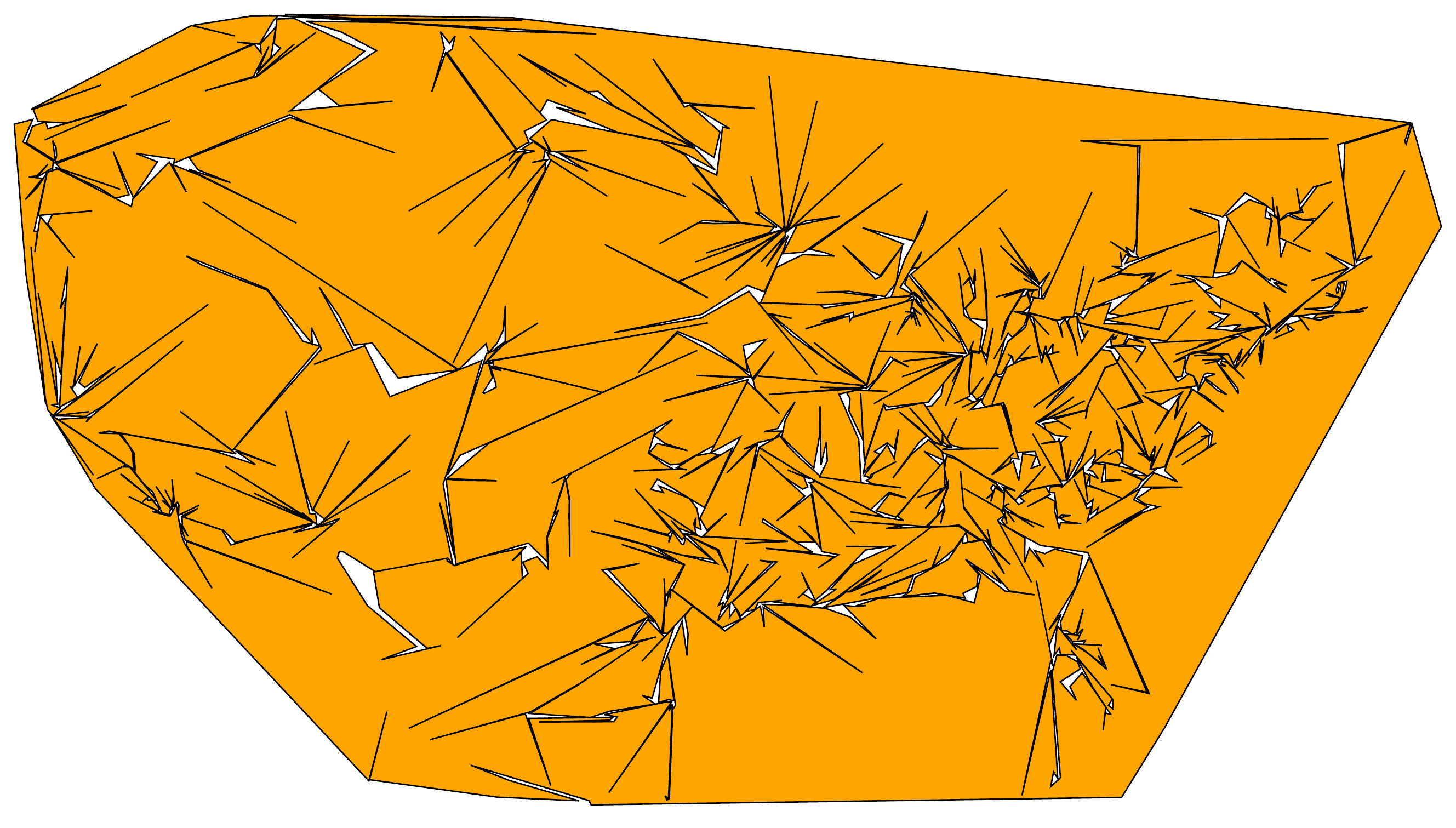} \hspace{.3cm}
    \includegraphics[width=0.47\textwidth]{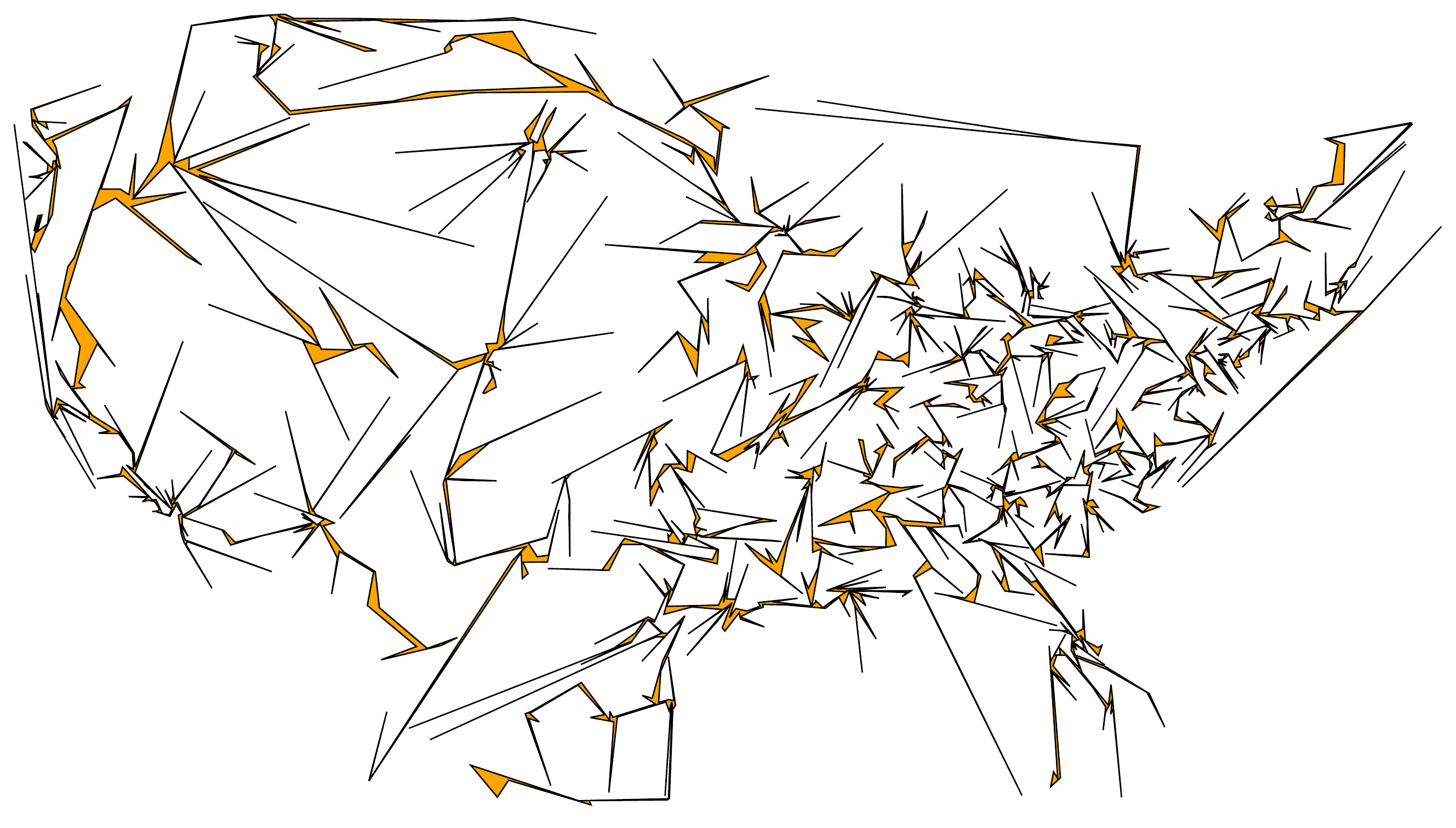} 
    \caption{\label{fig:minmax} Polygons of maximum and minimum area obtained for the \texttt{us-night-0002000} instance.}
\end{figure}

The strategy to find the polygon of small area is similar to the one described before to maximize the area and the polygons obtained often resemble each other except for the outermost edges (see Figure~\ref{fig:minmax}).
In fact, almost all lines of code are identical for both problems, except for fewer than 20 lines. The polygons are oriented clockwise for area minimization and counterclockwise for area maximization. This way, as signed areas are positive for clockwise orientation and negative for counterclockwise orientation, solving the problem to maximize the area addresses both objective functions.

The main difference between the two implementations is that instead of initiating the greedy algorithm with the convex hull, in the minimization version we start with a triangle and add vertices as we go. We use the same weight function as before, which allows us to add points that do not increase the area of the polygon by much while avoiding long edges. 
This procedure can be started with different triangles, which allows for better solutions by testing multiple triangles, but increases the running time significantly. Even though there are $\Theta(n^3)$ possible triangles with vertices in $P$, most of them do not seem appealing to start the greedy algorithm. Intuitively, we would like to start with a triangle of small area or perimeter. For example, one could choose to always start with the minimum perimeter triangle, but that choice would be very constraining and would eliminate the benefits of multiple starting configurations. As a compromise between 1 and $\Theta(n^3)$ triangles, we decided to use $O(n)$ possible starting triangles defined by a vertex $p_1$ as follows. We set $p_2$ to be the nearest neighbor of $p_1$ and $p_3$ to be the vertex that minimizes the perimeter of the starting triangle $t_0 = p_1p_2p_3$. This difference can be found on the \verb|poLYG.py| source code available on github on lines 63--75.

There are only two additional small technical changes that we need to make to the greedy algorithm. First, we need to add a test to make sure that the area of the polygon never changes sign by reversing the orientation of the polygon, as seen in lines 167--170 of \verb|geometer.py|.
Second, we only add triangles that increase the area in the greedy step, as seen in lines 106--110 of \verb|poLYG.py|. The reason is that there are too many triangles that would decrease the polygon area but are not valid since they cross other edges. Hence, testing all these intersections takes far too long. In theory, forbidding triangles that decrease the area could be problematic because if at any iteration an unconnected point lies inside the polygon, this point will never be added and no solution will be found. In practice this rarely happens.

	\section{Algorithm Engineering}
\label{sec:engineering}

In our fist implementation, we used a brute force approach to test whether each line segment intersects the existing edges of $P$, which takes $O(n)$ time. A naive implementation builds the final polygon $P$ by repeating, through $O(n)$ greedy steps, the test for all $O(n)$ edges $p_1p_2$ and $O(n)$ points $q$, testing intersection against $O(n)$ edges (see~\cite{ORo98} for the intersection test). This approach would take $O(n^4)$ time in both best and worst cases, which is far too slow for our purposes. Hence, we need to make it faster in practice, even if the worst case complexity does not improve.
This goal is achieved by (i) using binary heaps, (ii) using grids to speed up the intersection tests, and (iii) using grids to limit the set of points tested for each edge. Further improvements include a randomization step for the small instances and a divide-and-conquer strategy for very large instances (namely Mona Lisa, with 1 million points).


\subsection{Binary Heaps}

In order to improve performance, we want to perform as few intersection tests as possible. To do that, we use several minimum binary heaps. Each edge $p_{i}p_{i+1} \in P$ is associated to a heap that contains all points $q$ in $\barP$ (and possibly some points that are no longer in $\barP$). The priority of each point $q$ in the heap of $p_{i}p_{i+1}$ is set to $\weight(p_1,p_2,q)$. A higher-level minimum heap stores the minimum value of each binary heap.

We repeatedly extract the triangle $p_{i}p_{i+1}q$ of minimum weight and test if $p_iq$ and $qp_{i+1}$ intersect existing edges of $P$. If so, we repeat the procedure until a valid point $q$ is obtained, that is a point $q \in \barP$ such that $q$ may be inserted between $p_i$ and $p_{i+1}$ while maintaining simplicity. Then, we add the edges $p_iq$ and $qp_{i+1}$ creating the corresponding heaps. We remove the heap associated to the former edge $p_ip_{i+1}$. Notice that heap creation with $n$ elements takes $O(n)$ time while inserting an element and extracting the minimum value take $O(\log n)$ time. Since the number of intersection tests is significantly reduced, this approach yields a much faster running time in practice. 

\begin{figure}
    \centering
    \includegraphics[width=0.45\textwidth]{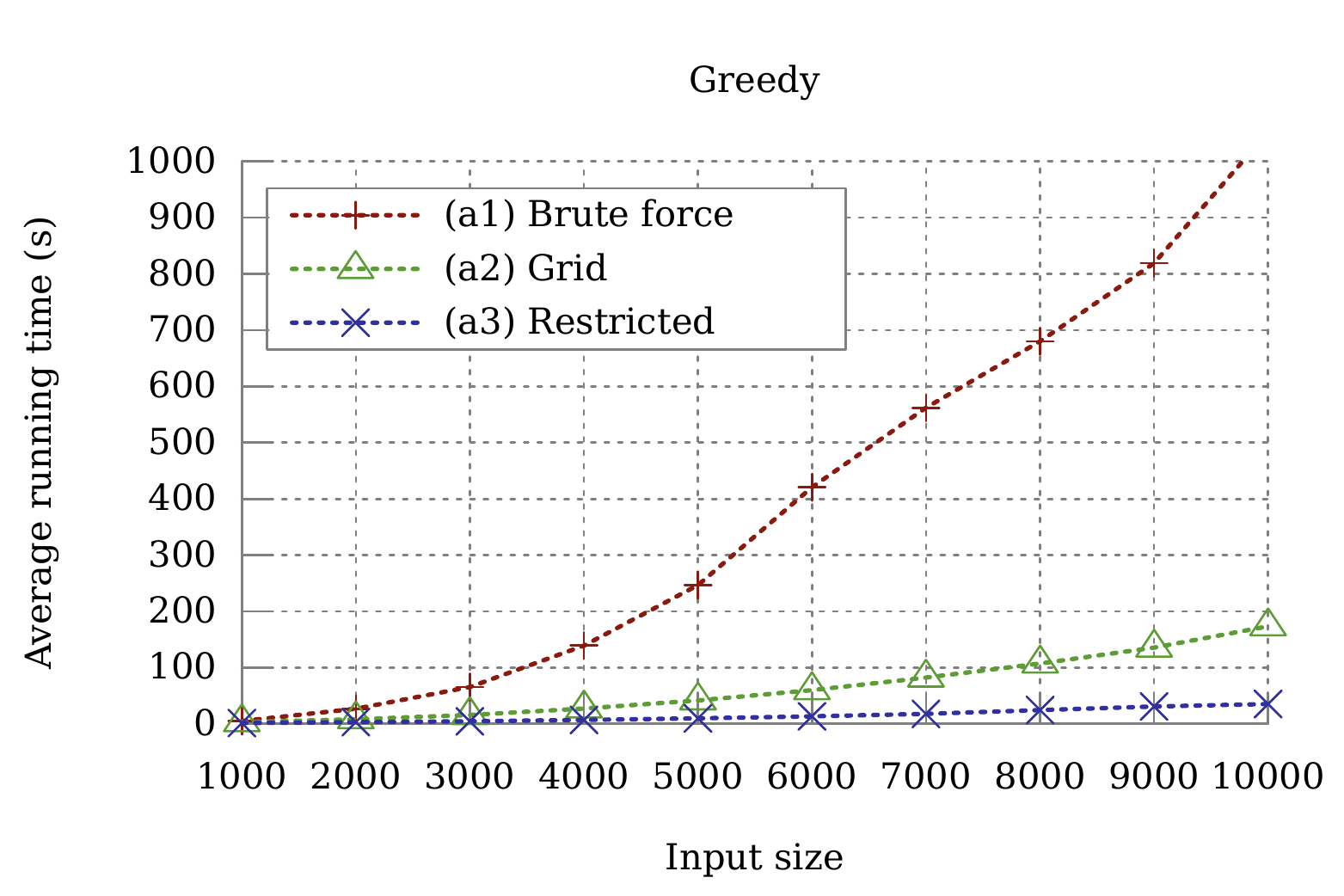}
    \hspace{.2cm}
    \includegraphics[width=0.45\textwidth]{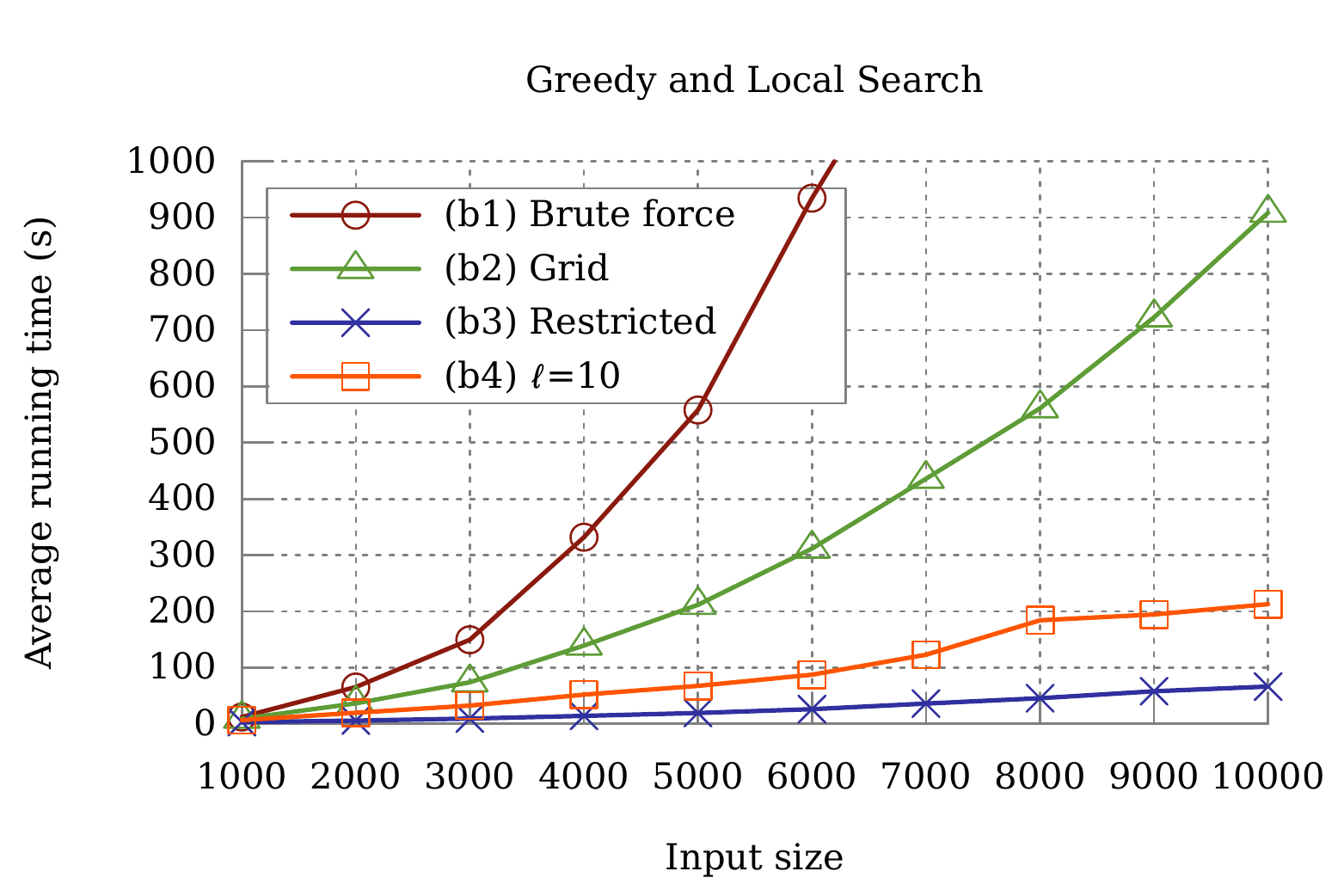}\\
    \hspace{1cm}(a)\hspace{0.45\textwidth}(b)
    \caption{Running time of (a) the greedy heuristic and (b) the total running time including local search as functions of the number of points. The values are an average over the $6$ different instances with the same number of points.}
    \label{fig:greedyplot}
\end{figure}

\begin{figure}
    \centering
    \includegraphics[height=5.1cm]{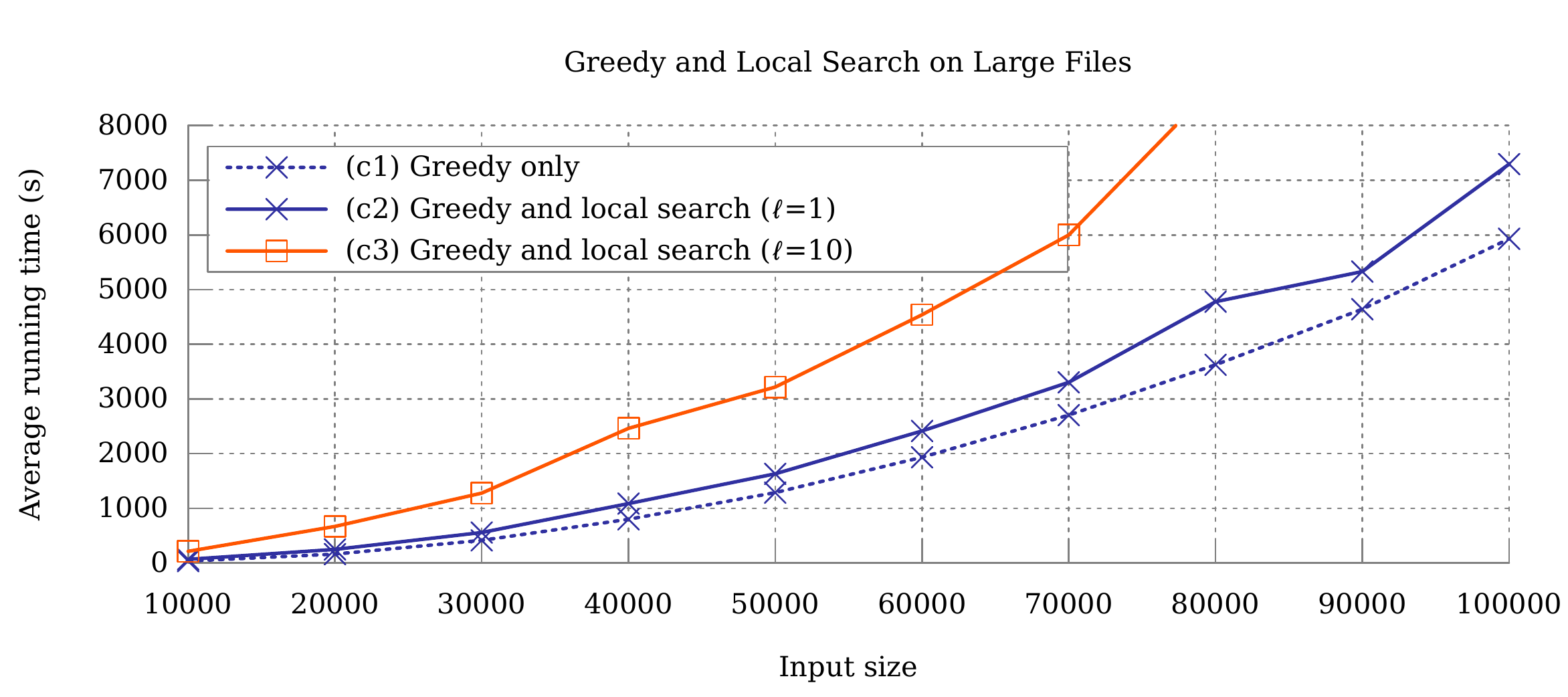}
    \caption{Running time of the greedy algorithm and the total running time including local search as a function of the number of points. The values are an average over the $6$ different instances with the same number of points.}
    \label{fig:largeplot}
\end{figure}{}

In a best-case scenario, the first triangle extracted from the heaps would always be accepted, which would lead to only one intersection test for each greedy iteration and a total running time of $O(n^2)$. Practical instances are not as good, but much closer to the best case than to the worst case of $O(n^4)$. 
The execution times for instances between $1,000$ and $10,000$ points is presented in Figure~\ref{fig:greedyplot}, plots~(a1) and~(b1). Running this code for $100,000$ points would take multiple days, but another serious impediment is the $O(n^2)$ space complexity of the algorithm. 
Due to the memory overhead of Python, an instance with $10,000$ points already uses over $5GB$ of RAM memory. Hence, an instance with $100,000$ points would require more than $500GB$ of memory. Next, we describe how to reduce the running time and later on, we will deal with the memory issue.

\subsection{Grid Intersection Test}

In order to speed up the algorithm, we perform a more efficient intersection test. We chose a very simple approach, with no worst-case guarantees. We partition the bounding box of the data using a square grid with the number of columns of roughly $(4n)^{1/4}$. If the points are uniformly distributed on a square, this gives $O(\sqrt{n})$ data points per grid cell. Each grid cell stores the list of edges that intersect that cell. Edges that intersect more than $4$ cells are kept on a separate list of \emph{long edges} instead. To test edge intersections with a line segment $pq$, we start by testing intersections with all long edges. If no intersection is found, then we trace the grid cells intersected by $pq$ and test intersections with the edges stored in each corresponding grid cell. 
The running time of our algorithm is presented in Figure~\ref{fig:greedyplot}, plots~(a2) and~(b2). Clearly, this approach is inefficient if there are too many long edges, but as explained before, our heuristic tries to avoid such edges.

\subsection{Restricted Candidates}

The same grid we described in the previous section is also used to partition the points of $S$ into grid cells. It seems rather unlikely that an edge $p_{i}p_{i+1} \in P$ will be connected to a point that is located many grid cells away from the cells intersected by that edge. For an integer $\kappa \geq 0$, we define the \emph{$\kappa$-neighborhood} of a grid cell $C$ as the set of non-empty grid cells within $L_\infty$ distance at most $\kappa$ of $C$. The $\kappa$-neighborhood of $C$ contains at most $(2\kappa+1)^2$ cells. We define the \emph{$\kappa$-neighborhood} of a segment $p_1p_2$ as the union of the $\kappa$-neighborhood of all cells intersected by $p_1p_2$.

Let $\kappa$ be a parameter of our algorithm. We limit the candidate triangles $p_ip_{i+1}q$ to triangles such that $q$ is in the $\kappa$-neighborhood of the segment $p_ip_{i+1}$. Setting $\kappa=2$ makes the algorithm run much faster  when compared to the previous case (which we call $\kappa = \infty$) with no noticeable impact on the quality of the solution. Furthermore, the memory requirement decreases significantly.
The results of the algorithm for $\kappa=2$ are presented in Figure~\ref{fig:greedyplot}, plots~(a3) and~(b3). 
The execution for larger instances with $\kappa=2$ is presented in Figure~\ref{fig:largeplot}.



\subsection{Randomization} \label{s:small}

With all improvements, the algorithm runs very fast on instances with fewer than $1000$ points. Since more than half of the challenge instances are in this category, it is important to spend additional effort to convert a longer computational time into improved solutions. For this purpose, we used a very simple modification of the greedy algorithm from Section~\ref{s:greedy} followed by the same local search optimization.

We use a Gaussian random variable $\gauss(\sigma)$ of standard deviation $\sigma$ to obtain the following randomized weight function, where the original weight function is multiplied by $1 + |\gauss(\sigma)|$. 
\[\weight(p_1,p_2,q) = (1 + |\gauss(\sigma)|) \cdot \big(\area(p_1p_2q) + \alpha(\|qp_1\| + \|qp_2\| - \|p_1p_2\|)\big)\]

\begin{figure}[b]
    \centering
    \includegraphics[height=0.32\textwidth]{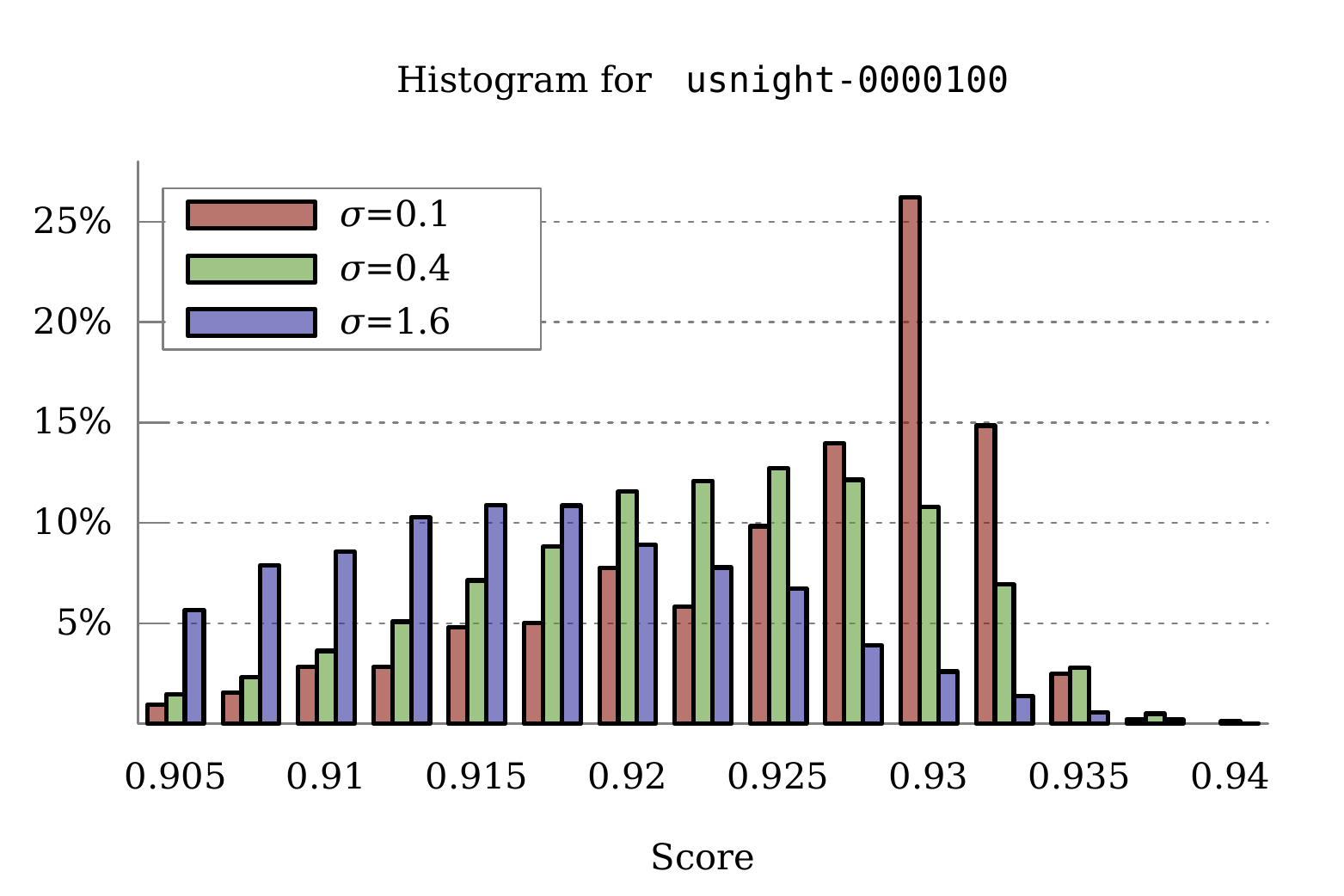}
    \hspace{.2cm}
    \includegraphics[height=0.32\textwidth]{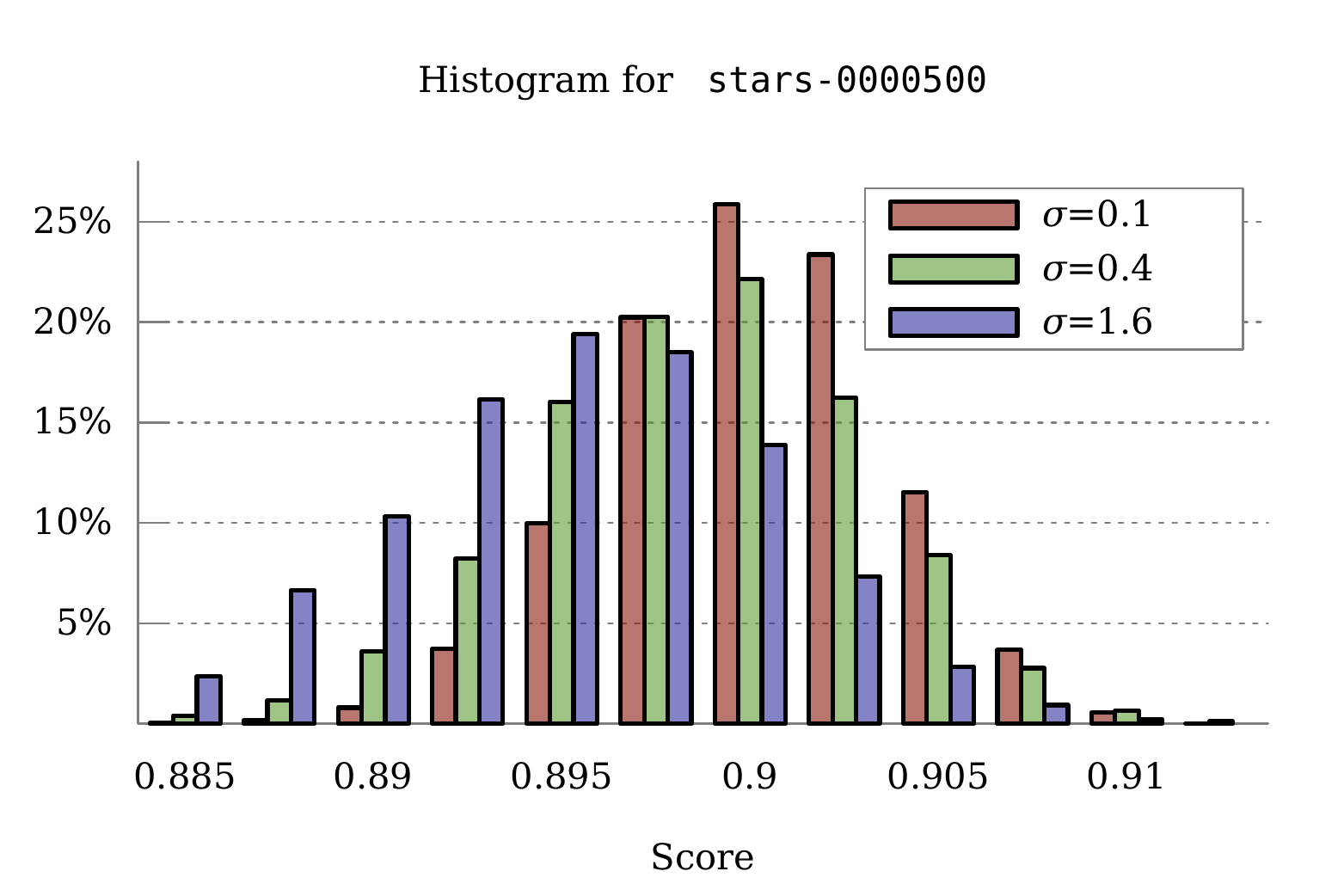}
    \caption{Histogram of the distribution of scores of solutions found for files \texttt{usnight-0000100} (100 points) and \texttt{stars-0000500} (500 points) with different values of $\sigma$.}
    \label{fig:hist}
\end{figure}

The weight  is calculated only once per triangle and kept throughout the execution of the greedy heuristic. Furthermore, we set $\kappa=\infty$ for instances with at most $100$ points, and keep $\kappa=2$ otherwise.

In Figure~\ref{fig:hist}, we show two histograms of the scores of the solutions obtained with different values of $\sigma$. The histograms were obtained by rounding down the scores to a multiple of $0.0025$ among $5000$ solutions for each instance and value of $\sigma$. Some lower scores are not shown. We used $\ell=1$ for both files, $\kappa=\infty$ for the files with at most $100$ points and $\kappa = 2$ for $500$ points. The time to obtain these $5000$ solutions is around $6$ minutes for $100$ points and $30$ minutes for $500$ points. 

\begin{figure}
    \centering
    \includegraphics[height=4.8cm]{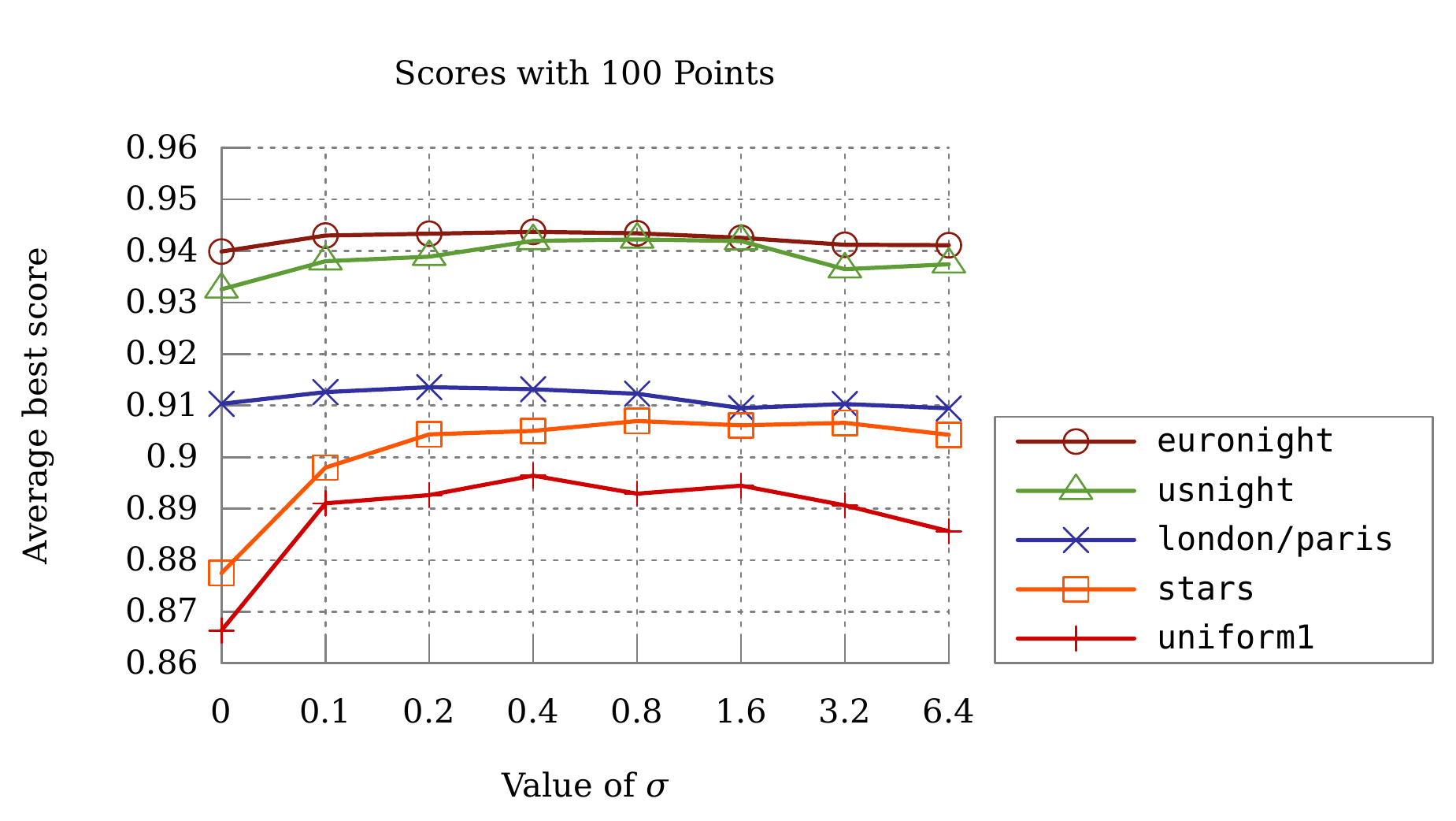}
    \includegraphics[height=4.8cm]{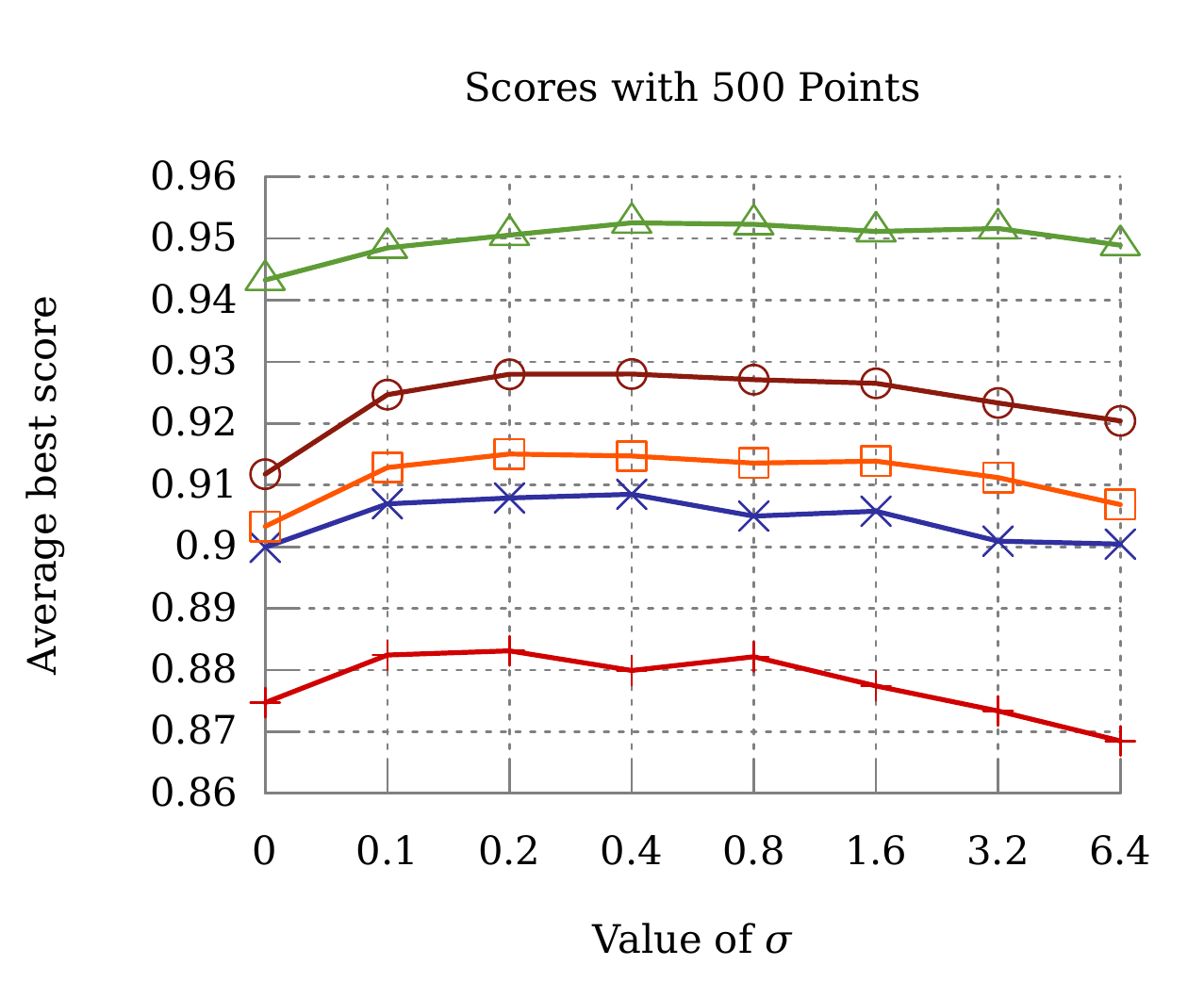}
    \caption{Score as a function of $\sigma$ for different instances with 100 and 500 points.}
    \label{fig:sigmaplot}
\end{figure}

Using the same parameters, the best scores for different values of $\sigma$ and several files with $100$ and $500$ points is represented in Figure~\ref{fig:sigmaplot}. The plot shows that the best values of $\sigma$ are generally between $0.2$ and $0.8$ and that using multiple values of $\sigma$ for each file while keeping the best solution found will give a slightly better score.

Setting $\sigma=0.5$ we obtain an average of $0.918$ among the $84$ files with up to $100$ points and $0.911$ among the $48$ files with $200$ to $900$ points. The solutions submitted to the challenge have slightly better average scores of $0.921$ and $0.913$, respectively. This small improvement can be obtained by trying multiple parameters and keeping the best score for each file.

\subsection{Mona Lisa} \label{s:monalisa}

Since the algorithm we presented would take several weeks for one million points, we use a divide and conquer strategy for the \texttt{mona-lisa-1000000} file.

We partition the points from $S$ into a regular square grid of $32 \times 32$ cells. This way each grid cell contains approximately 1000 points on average. We then separately execute the algorithm to the points contained in each grid cell.
At the end of the execution, we have 1024 simple polygons that we merge into one single simple polygon as follows.

First, for each cell, and for each of the (up to 4) adjacent cells we compute the best way to merge the two polygons associated to those cells.
Let $P_1$ and $P_2$ be the two polygons that we desire to merge. Then, for each edge $e_i$ in $P_1$ and each edge $e_j$ in $P_2$, we compute the quadrilateral $Q_{ij}$ supported by the edges $e_i$ and $e_j$. Among all quadrilaterals $Q_{ij}$ that intersects neither $P_1$ nor $P_2$, we select the one with the largest (or smallest depending on the objective) area. We call this quadrilateral $Q_{ij}$ a \emph{bridge}. This construction guarantees that $P_1 \cup Q_{ij} \cup P_2$ is a simple polygon.

Note that it is possible for two bridges to intersect one another. We fix this issue by forbidding the new bridges to intersect the ones that have been computed previously.
We now have to select which bridges should be used to connect all the 1024 polygons into a single one.
If we consider all cells of the grid as vertices of a graph $G$ and all bridges as edges of $G$, connecting our 1024 polygons is equivalent to computing a minimum spanning tree in $G$. We compute this spanning tree using Prim's algorithm. 

	\section{Results}
\label{sec:results}


Our algorithm is controlled by four parameters: $\alpha$ for the weight penalty of the long edges in the greedy routine, $\ell$ for the maximum length of the path investigated in the local search,  $\kappa$ for the size of the neighborhood of an edge in which we are searching for new vertices and at last, $\sigma$ for the randomization. In the software we made available, $1/\alpha,\ell,\kappa,\sigma$ are respectively called \texttt{pen}, \texttt{hops}, \texttt{hood}, and \texttt{sigma}.

For small instances (at most $1000$ points), the speed of the algorithm allows us to run a very large number of  experiments in order to keep the best solution found.
The parameter $\kappa$ is fixed at $+\infty$, $\ell$ is equal to $10$ and we used different values of $\alpha$ and $\sigma$, keeping the best score.

For medium instances ($2,000$ to $10,000$ points), we use $\kappa=2$ to accelerate the algorithm and try different values of $\alpha$. We avoid using randomization ($\sigma \neq 0$) as each execution is too slow to allow for a sufficiently large number of executions.

The average score obtained for the instances with $10,000$ to $100,000$ points using only the greedy algorithm, $\kappa =2$, and $\alpha=1/90$ is $0.892$. Using local search with $\ell=1$ the average score improves to $0.910$. Setting $\ell$ to $10$ increases the average score to $0.912$. Increasing $\ell$ beyond $10$ made no difference to the solutions obtained. In the challenge, the average score we got for the same instances was close to $0.915$. The small improvement can be achieved by using multiple values of $\alpha$, increasing $\ell$, and using randomization with multiple values of $\sigma$. The times of computation of a single solution are shown in Figure~\ref{fig:largeplot}.


We present some of the results obtained for some of instances of the challenge CG:SHOP 2019 in Table~\ref{tab:results}.

\begin{table}
\begin{tabular}{|R{1cm}|C{1.5cm}|C{1.5cm}|C{1.5cm}|C{1.5cm}|C{1.5cm}|C{1.5cm}|}
\hline  & \multicolumn{2}{c|}{euro-night} & \multicolumn{2}{c|}{us-night} & \multicolumn{2}{c|}{uniform1} \\
size & min score & max score & min score & max score & min score &  max score   \\
\hline
10 &  0.312 & 0.893 &  0.251 & 0.898 &  0.352 & 0.884\\
\hline 
20 &  0.140 & 0.910 & 0.114 & 0.940 &  0.214 & 0.864\\
\hline 
30 &  0.126 & 0.923 &  0.081 & 0.930 &  0.176 & 0.920\\
\hline 
40 &  0.187 & 0.944 & 0.075 & 0.945 &  0.161 & 0.917\\
\hline 
50 &  0.113 & 0.923 &  0.063 & 0.959 &  0.091 & 0.925 \\
\hline 
70 &  0.085 & 0.938 & 0.078 & 0.952 &  0.117 & 0.921\\
\hline 
90  & 0.075 & 0.947  & 0.097 & 0.933 & 0.137 &  0.898\\
\hline 
200  & 0.085 & 0.929 &  0.073 & 0.941  & 0.123 & 0.895 \\
\hline 
400 &  0.079 & 0.934 & 0.061 & 0.950 & 0.135 &  0.883\\
\hline 
600 &  0.071 & 0.936 &  0.059 & 0.948 & 0.121 & 0.890\\
\hline 
800 &  0.068 & 0.942 & 0.049 & 0.953 &  0.130 & 0.876\\
\hline 
1000 &  0.065 & 0.943 &  0.049 & 0.955 & 0.130 & 0.871\\
\hline 
3000 &  0.061 & 0.942 &  0.043 & 0.957 &  0.122 &  0.880\\
\hline 
5000 &  0.055 & 0.947 &  0.041 & 0.959 &  0.125 & 0.878\\
\hline 
7000 & 0.052 & 0.949 &  0.038 & 0.962 &  0.127 & 0.873\\
\hline 
9000 &  0.050 & 0.949 &  0.036 & 0.965  &  0.127 & 0.872\\
\hline 
20000 &  0.046 & 0.955 & 0.032 & 0.968 &  0.126 & 0.874\\
\hline 
40000 &  0.041 & 0.959 &  0.029 & 0.972 &  0.124 & 0.876\\
\hline 
60000 &  0.038 & 0.962 & 0.027 & 0.973 &  0.125 & 0.876\\
\hline 
80000 &  0.038 & 0.964 &  0.025 & 0.976 &  0.124 & 0.877\\
\hline 
100000 &  0.035 & 0.965 &  0.025 & 0.975  & 0.124 & 0.877\\
 \hline 
\end{tabular}
\caption{\label{tab:results} The results of our heuristics for several instances on the 2019 challenge.}
\end{table}

	\section{Concluding Remarks}
\label{sec:outlook}

The greedy algorithm that we use in the first phase is highly dependent on the weight function chosen. Figure~\ref{fig:nopen} shows the huge difference between two results obtained with a weight function with and without a penalty for long edges. We tested different penalties and different values of $\alpha$, but these investigations have not been done in an exhaustive manner. It is likely that better results may be obtained by customizing the weight function for different types of files, depending on specific characteristics of the instances. Even without changing the expression of the weight function, a deeper investigation on the dependency on the value of $\alpha$ is desirable.

The local search step that we use in the second phase can also be improved with more sophisticated tools such as meta-heuristics and heuristics developed for the TSP problem, as it has been done by other teams during the challenge. The combination of our greedy heuristics with the more advanced local optimization could provide better results than the ones that we obtained independently.

\section{Acknowledgments}

Loïc Crombez has been sponsored by the French government research program ``Investissements d'Avenir'' through the IDEX-ISITE initiative 16-IDEX-0001 (CAP 20-25). Guilherme D. da Fonseca and Yan Gerard are supported by the French ANR PRC grant ADDS (ANR-19-CE48-0005).

We would like to thank Hélène Toussaint, Raphaël Amato, Boris Lonjon, and William Guyot-Lénat from the LIMOS HPC cluster, whose computational resources were extremely useful for the competition. We would also like to thank the challenge organizers and other competitors for their time, feedback, and making this whole event possible.


\bibliographystyle{abbrv}
\bibliography{references}

\end{document}